# A framework for developing a knowledge management platform.

Marie Lisandra Zepeda Mendoza[a], Sonali Agarwal[b], James A. Blackshaw[c], Vanesa Bol[d], Audrey Fazzi[d], Filippo Fiorini[e], Amy Louise Foreman[c], Nancy George[c], Brett R. Johnson[f], Brian Martin[g], Dave McComb[h], Euphemia Mutasa-Gottgens[c], Helen Parkinson[c], Martin Romacker[i], Rolf Russell[j], Valérien Ségard[d], Shawn Zheng Kai Tan[c,k], Wei Kheng Teh[c], F. P. Winstanley[l], Benedict Wong[j], and Adrian M. Smith[m]

[a] *Machine Intelligence, AI & Digital Research, Novo Nordisk Research Centre, Oxford, UK;* [b] *GSK Global Capability Center, Bagmane Capital, Mahadevapura, Bengaluru, Karnataka;* [c] *EMBL-EBI, Wellcome Genome Campus, Hinxton, UK;* [d] *GSK, Vaccines R&D, Rue de l'Institut 89, Rixensart;* [e] *GSK, Via Fiorentina , 53100 Siena, Italy;* [f] *Global Genes, 28 Argonaut, Suite 150, Aliso Vejo, CA;* [g] *AbbVie, 1 N Waukegan Road, AP4/B, North Chicago, IL, United States of America;* [h] *Semantic Arts, 123 N College Avenue; Suite 218; Fort Collins, CO;* [i] *F.Hoffmann-La Roche, Grenzacherstrasse 124, CH-4070 Basel;* [j] *TXI, 10 S Riverside Plaza Suite 875, Chicago, IL;* [k] *Data Representation, Novo Nordisk A/S, Måløv, Denmark;* [l] *Semantic Arts, Kilcreggan, Scotland;* [m] *Unilever R&D, Colworth Science Park, Sharnbrook, Bedford, UK.*

**Abstract**

Knowledge management (KM) involves collecting, organizing, storing, and disseminating information to improve decision-making, innovation, and performance. Implementing KM at scale has become essential for organizations to effectively leverage vast accessible data. This paper is a compilation of concepts that emerged from KM workshops hosted by EMBL-EBI, attended by SMEs and industry. We provide guidance on envisioning, executing, evaluating, and evolving knowledge management platforms. We emphasize essential considerations such as setting knowledge domain boundaries and measuring success, as well as the importance of making knowledge accessible for downstream applications and non-computational users and highlights necessary personal and organizational skills for success. We stress the importance of collaboration and the need for convergence on shared principles and commitment to provide or seek resources to advance KM. The community is invited to join the journey of KM and contribute to the advancement of the field by applying and improving on the guidelines described.

## 1. What is knowledge management and why do we need it?

Knowledge management (KM) is the set of techniques, processes, tools, and strategy by which data are converted into knowledge that can be stored, shared, analyzed, expanded, and turned into insights and impact (Senge, 2006). Systems at scale are needed to handle large amounts of information for rapid and unbiased analysis of incoming and existing datasets. By defining relationships between concepts agreed upon by subject matter experts (SMEs), a KM platform (KMP) provides the framework to ingest, curate, contextualize, harmonize, integrate, and formulate hypotheses from all available data. Accompanying metadata and information on where it has been sourced provides secure versioning and provenance. Such information may be particularly difficult to gather for historic data, however, historic data that is still relevant can be used as the foundation of a KMP if determined by the SME. Given the immediate availability of historic data, it can be a challenge to select what initial knowledge to capture and how to model it.

It is critical to select appropriate ontologies to model the data by defining the relevant entities and how they are linked. One should consider the quality of the ontology, as well as how many of the terms from the chosen ontology map to other available ontologies (e.g., in open resources like Biomappings or OxO) (EMBL-EBI, 2023; Hoyt et al., 2023). A different strategy is required for real-time capture of data that is currently being generated, or that will be generated under a defined timeline. Depending on the differences between new and existing data sources, new processing code may also need to be integrated. Capturing experiential knowledge from SMEs is also valuable as this knowledge is at risk of being lost when the SMEs retire or redeploy. However, capturing tacit knowledge requires different processes and toolsets, such as interviews, feedback reports, log reports, etc. Therefore, it is important to consider how to associate domains with users instead of creators, as cross-domain data usage becomes more prevalent. Sometimes this involves forward planning to ensure that permission for data capture and use, such as activity tracking, is gained in advance.

The necessary aspects for data storage, use, and sharing are now well understood and practiced in academia and industry, especially through the frequently discussed standards of Findability, Accessibility, Interoperability, Reusability (FAIR) (Wilkinson et al., 2016); and Transparency, Responsibility, User focus, Sustainability, and Technology (TRUST) (Lin et al., 2020). These common data standards, in combination with data storage technologies that have passed internal safety-risk evaluations, allow for stewardship and KM practices to return increased value to data and analytic practices. However, the use of these for knowledge creation and engineering through common practices is less well defined. There is a need for similar shared practices, methods, and standards to be universally agreed for effective KM.

Common KM standards in curation and analytics processes, and robust governance operations that are shared across public and private sectors, will enable 1) sharing of unbiased and interpretable knowledge for better decision making; 2) faster insights, leading to increased productivity; 3) the extraction of new knowledge for increasingly innovative products; and 4) reduced barriers, breaking down of silos, and increasing return on investment for academia and industry.

**2. How can we manage knowledge?**

The definition of a data model for KM involves careful evaluation of the sources of data from which knowledge is taken, their redundancy, update status, and noise. Once designed and tested, the model may require re-evaluation of its complexity and if necessary, be redesigned. Furthermore, there should be methods in place for securing provenance as new data is ingested and modified versions of the platform are archived (e.g., for reproducibility of project analytics or to comply with legal requirements).

A centralized KMP can be seen as a knowledge base meant to satisfy users, at least to a minimum specified level, or as a platform that allows the extraction of knowledge needed for the user to create their own task-specific knowledge bases. In principle, both scenarios can be obtained from the same single platform design. A central KMP needs to be easy to access, navigate, query, and pull data from – both directly via a user interface and computationally via e.g., an application program interface (API). SMEs can use the already harmonized and annotated knowledge to build their own models which they can enrich with their own selected data and apply any manual curation steps without restrictions, according to the needs of their field. Ideally, each spun-off knowledge base should be able to choose their own technology provider, although utilization of a single technology across the organization is advised to enable future integration.

Usually, the teams developing KMPs in an industrial context are either centralized with a hierarchical structure or organically evolved and set up in a more distributed manner in different sections of the organization. A centralized approach to KM within a very diverse organization can create bottlenecks and reduce flexibility. Federating contributors may address these challenges as it enables autonomy and independence of each part of the organization. However, federated KM could require a higher integration effort and be under higher risk of duplicating efforts, particularly if there is not a unified vision. A federated approach is most effective if robust data stewardship is in place and can produce a platform with a broader scope while retaining a cohesive and unified direction which meets user needs.

### *2.1 Definition of knowledge domains*

There can be different approaches for building a KMP, depending on the underlying technology and data sources, as well as the number and overlapping nature of domains. Knowledge domains are specialized fields where experts have a level of understanding and experience. It is important to define the knowledge domain that will be part of a KMP and identify which concepts are truly core and which belong to subdomains aligned to a specific use case or choice of interpretation. There are infinite possible areas of knowledge with different vocabularies and ontologies, making it challenging to create an exhaustive list of knowledge domains. The complexity of this task is illustrated by the CYC project's objective to codify human common sense, which has been going on since 1984 and estimated to require up to 3000 person-years of effort (Lenat et al., 1986).

Customer-oriented and use-case driven frameworks can be used to identify the boundaries of a knowledge domain. These approaches involve identifying business needs and collaborating across data silos, rather than relying solely on SMEs. However, it is important for all parties involved in the KM project to align on its purpose. Interviews with upper management and key stakeholders help to determine broad and specific terms and concepts related to the topic, as well as data sources and user stories. With this information, the domain boundary can be formalized by customizing existing controlled vocabularies or modifying an existing ontology to capture key terms and relationships within the domain and validating the representation of the domain with key stakeholders.

*2.2 Scope, scale, and speed as factors influencing the domain boundaries.*

It is difficult to define the limit of information that should be included in a KMP and to identify when areas of the model are not actively useful. Technical and structural limitations and the frequency of modifying, sharing, and accessing the platform can help determine when the model needs to reconsider its scope and scale. Scope refers to the variety of data; it defines how broad the platform is intended to be in terms of users and sources of content, while scale refers to the volume and depth of data and how complex it will be in terms of maintenance, presentation, and usage. Speed is tied to both scope and scale and is influenced by the resources allocated for the platform development. When defining the scale of a KMP, the depth of annotation for each of the concepts in scope needs to be defined, as well as storage and visualization of such concepts and related annotations. The interface should provide intuitive mechanisms for data presentation and exploration while having encoded specific limits where appropriate to protect platform stability and speed.

It is tempting to try to satisfy every potential use case from the outset. However, Agile project management, where an initial minimum viable product (MVP) has been defined, is often the best way of getting something robust in place, showcasing that it can provide value and thus justifying the expenditure of more resources to keep supporting its expansion (broadening its scope) and maintenance (facilitating future scalability). An Agile way of working, where first an MVP is defined, and program increments deliver products with increased complexity and satisfying more use cases, is a good way to balance speed and expectations. Regardless of the speed, best practices should be implemented from the beginning and never compromised for the sake of saving time. Early, well-defined platform specifications can prevent uncalculated, rapid initial decisions from locking in bad choices that will be expensive to fix later. Limiting knowledge to the boundaries of a chosen scope may of course limit the discovery of new insights, which is a key benefit of a KMP, therefore it is important to carefully consider the trade-off between scope and speed in the development, quality and completeness of a KMP.

**3. What technical aspects need be considered to develop a data model for a KMP?**

There are two main technical choices to make to develop a KMP: 1) the data storage technology (e.g., knowledge graph, data lake, warehouse, fabric), and 2) the ontologies to model the data. Just like the specific storage technology requires a specific set of data engineering and infrastructure skills, ontologies also require specific technical skills. Ontologies are difficult to build, maintain and edit, as they require theoretical and technical engineering knowledge to practically handle them, as well as domain knowledge of the area being modelled. It is not advised to create new ontologies, but to reuse as much as possible reference ontologies (e.g., OBO Foundry ontologies (Jackson et al., 2021)) and extend them if needed. Sometimes, this ontology extension can be done as a community effort (Matentzoglu, Goutte-Gattat, et al., 2022; Ong & He, 2016).

Tools that allow easy visualization, editing, dynamic imports, version control, data export, etc. (Côté et al., 2010; Matentzoglu, Goutte-Gattat, et al., 2022; Mungall et al., 2023), are crucial to the management and access of the ontologies. This will allow you to make changes without affecting downstream users and will also let consumers of your KMP choose when they update their ontology or models, where new versions might cause issues to their pipelines. There are tools to modify existing ontologies that are based on established ontology description methods (e.g. OWL (RDF/XML) (OWL Working Group, 2013), OBO (Tirmizi et al., 2011)), and tools to interoperate between such different descriptions (e.g. ROBOT (Jackson et al., 2019)).

An ontology is not a static artifact, but it evolves together with the domain being modelled. Ontologies should drive and be driven by data capture, for example, through the utilization of metadata templates to model samples and results or through informing discussions with SMEs when capturing tacit knowledge. Controlled vocabularies should be used in partnership with ontologies and metadata templates to ensure that e.g., synonyms and alternate spellings do not create false issues with data quality. Good ontology management also requires constant curation activities from curators that have knowledge of both the domain and ontology modelling, ideally working closely with SMEs and semantic developers. Automated processes, such as quality control, can make curation much more efficient, lowering the overhead in resources required and can help in delivering a more internally consistent ontology (Matentzoglu, Goutte-Gattat, et al., 2022).

*3.1 How to make data from a KMP accessible for downstream applications and non-computational users?*

It is important for an organization to have well-defined processes, roles, and responsibilities to ensure the flexibility of knowledge collection and sharing to downstream applications (e.g., through APIs). Tools such as Swagger/OpenAPI (Swagger, 2023a, 2023b), AWS software development kits (SDK) (Amazon Web Services, n.d.), and other command line applications to build simple scripting can precede more complex applications (Cherry, 2022; Mulesoft, 2023). However, creating software that targets developers is itself a sub-discipline. Building programming interfaces to a KMP, including those based on query languages, is part of this discipline and patterns around API and SDK development apply here. Standard query front-end tooling can also be used, including low-level relational database tools (e.g. Toad or DataGrip (DataGrip, 2023; Scalzo & Hotka, 2009)) and those that allow graphical query design requiring a low-level understanding of the data model. There are also tools like Dash, Spotfire, or Tableau (Ahlberg, 1996; Hossain, 2019; Tableau, 2023) allow rapid application development of reduced-scope tasks. Alternatively, web applications can be developed, allowing an easier access to a large audience.

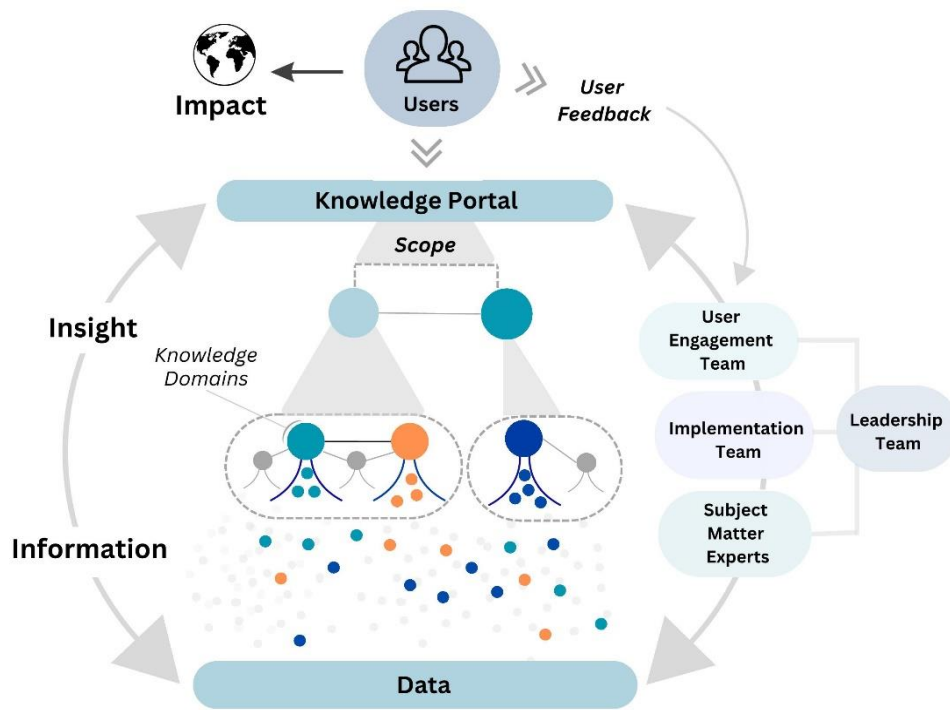

**FIGURE 1** | Knowledge is the understanding of connections between data points; and insight is when that knowledge is contextualized, and applications are found for it - ultimately leading to impact. Importantly, insights lead to activities that generate more data, that can turn into more knowledge. Knowledge can be organized into defined domains and subdomains, with ontologies unifying knowledge from different sources and domains, making them interoperable and harmonized into a single KMP. As the platform matures, it should allow for easy extraction of task-specific data, to allow for the making of task-specific data models or platforms. Such spin-off platforms can be short-lived or can be enriched with more data and then be ingested back into the original KMP. This approach seamlessly follows the way that knowledge itself evolves and feeds-back. A KMP can only be sustained through the constant committed work of a leadership team who fosters a culture of using the KMP and provides funding for the implementation team. The implementation team develop the platform following input from subject matter experts, and the advice of the user engagement team; which is in turn highly connected to the users to build a backlog of implementation tasks that can be tractable, actionable and measurable.

## 4. What personal and organizational skills are needed for implementing a KMP?

To effectively implement and accelerate value creation through a KMP, an organization needs the right mindset, skills, and vision to enable the cultural transformation required for everyone to embrace the use of the KMP across the value chain. Organizational culture is the set of values, expectations and practices to which employees are exposed. How participants are trained and how they observe others creates a critical atmosphere for the KMP to extend throughout an enterprise. When engaging a communication team to support the culture changes that surround a KMP, it is important to display early adopting program wins to all stakeholders throughout the organization and making sure those in domains not initially participating in pilots are involved and do not feel ignored.

A successful KMP that provides significant business impact requires stakeholders with the appropriate organisational influence, budget, mindset, and skillset to drive it forward. Stakeholders are those who have invested interests in the outcome of the project. The development of a KMP may include a different number of stakeholders, depending on the platform's scope and scale, but there are three **key stakeholders** needed for the development of any KMP:

> **Leadership team (LT):** Stakeholders who provide direction, scope, and funding for the KMP. An ideal LT for a KM project should have skills in managing budgets and people, influencing other leaders, and the authority to roll out the project. The LT should have realistic expectations and proper knowledge of the KMP's purpose by taking clear ownership of the project, agreeing on domains, goals, and milestones, and ensuring the project aligns with the overall organizational strategy. The LT should commit to the project for the long term and be prepared to remove blockers to progress, as they can align their vision and strategy to the long-term potential of the KMP.

**Implementation team:** Stakeholders who design and build the project within the given scope and budget. Planning, discussing, and sharing ideas are vital to align expectations, but eventually implementation needs to be done. Implementation stakeholders are responsible for turning tasks into deliverables. They have the required technical expertise to complete tasks with the quality and timeliness required by the KM team and the organization. They are likely SMEs and have the necessary skillset for the part of the system they are contributing to, with credibility within the organization and collaborations across the teams involved in the KM project. The mindset of the implementer should allow them to see and believe in the overall KM project vision, providing ideas while keeping an informed scepticism to challenge assumptions and ensure the implementation remains aligned with the overall goals and scope of the project to avoid feature creep. There would usually be an Agile-minded project leader or leaders to bring this multi-disciplinary team together in an effective manner.

**Users and user engagement team:** The users who extract knowledge from the platform, and the data providers who supply data inputs. The potential impact of a KMP is related to the number of users and use cases. Engagement with end-users is vital for the platform to be accepted and useful. The project team may need to perform interviews, gather feedback, and organize onboarding activities with them, and could group them into alpha and beta testers and final end-users. Ideal users provide high-quality requirements and give honest and actionable feedback that relates to the goals of the KM system. They are keen to use KM tools or are open to new approaches, patient with the development of the tools, and well-informed and curious regarding what it takes to build a KMP and understand the data it models.

## 5. Measurements of the success and maturity of KM in an organisation

Metrics are critical to assess the effectiveness and efficiency of their KM practices, track progress over time, make informed decisions, demonstrate the value of KM to stakeholders, and advance achievement of overall business objectives. Metrics to measure the success of a KMP initiative should align with the strategic priorities of the team, function, and organization. Clear metrics should be defined in line with stakeholders needs and the expected impact, such as productivity increase, user engagement, cost and time saving, intellectual property, and return of investment. There are many different metrics that can be used in the context of KM, including both quantitative and qualitative methods. The decision on which metrics to employ depends on overall business objectives, stakeholder needs, industry benchmarks, etc. The metrics should represent the aspect of interest without causing too much burden while being easy to implement. Examples of metrics for a KMP may include user engagement measured using system access, productivity increase measured through response time, and cost impact measured using reduction in the number of customer service tickets raised per quarter. Establishing a baseline is important before attempting to measure the success of any effort to improve KM practices. By collecting and analysing data on key performance indicators related to KM, organizations can identify areas for improvement and make informed decisions about how to optimize their efforts.

**Conclusion**

*Knowledge management as a shared cross-organizational, and cross-industry, journey*

This white paper is the outcome of a compilation of concepts and ideas derived from multiple workshops on KM organized by the EMBL-EBI Industry Program, edited by SMEs at the EMBL-EBI and participating industries. The field of KM is rapidly evolving and being applied to a variety of fields, and there is no single source of truth. The goal of this white paper is to encourage collaboration and sharing of approaches, methods, and tools, and to begin the discussion towards standardization of approaches – which together will enable the community to leverage the vast amount of knowledge that is already available and guide the production, organization, and impact of future knowledge. This is an activity that has applicability across all industries and need not be contained within Life Sciences.

Looking back at all the years of KM, we believe there has never been a true lack of resources. NLP techniques, various database technologies, ontology management systems, etc., have always been there. Despite this, the community is just as disconnected today as it was before the emergence of recent technologies designed to harmonize it. We believe the current challenge in KM is the sponsorship and ownership of activities, not only between domains, but between academia and pharma. Regarding the development of an inhouse KMP, we advise to design a proof of concept and seek funding for it. This will act as a tool for lobbying funding sources to explore implementation models.

Some mindset and cultural actions can be taken today by departments undertaking KM to start walking in the same direction along a cross-organizational and cross-industry KM journey. For example, designing data sources and catalogues with automation and self-service in mind, to reduce the need for manual work and thus harmonize processes and make their sharing easier. Instead of making catalogues of catalogues, it would be a better approach to make solid processes for automatically assembled data catalogues as federated objects. We believe it is the convergence on shared principles to disseminate data, rather than the data list itself, that will advance the field.

The EBI's effort at assembling an academic-industrial community around the vision of common KM practices and sharing of resources, is one of the first steps to pave the way of the shared journey we call for. Future activities could include the developing of a public/private bridging ontology that integrates existing ontologies in academia and industry with the aim of forming an interdisciplinary shared knowledge representation. This should build on prior and ongoing work, including the SSSOM standard (Matentzoglu, Balhoff, et al., 2022) and OBO Foundry's work on shared ontologies across academia (Harrow et al., 2019; Hoyt et al., 2023; Mungall et al., 2020).

The idea of a community contributing to the advancement of the KM field is not a short-term or one-time-shot activity, but a long-term process that requires agreement on principles and commitment to provide or seek resources - like has been achieved for the management and sharing of data. Different communities have a number of resources (data, tools, practices) that they play by, but it is through opening up to sharing best practises as well as code and tools, that the journey can be more bearable. As a start, the community is invited to join the journey of KM and contribute to the advancement of the field by applying and improving on the guidelines described here.

## Acknowledgements

We would sincerely like to thank the speakers of the two EMBL-EBI Industry Programme workshops in Knowledge Management held in Cambridge, UK in May and July 2022 for thought provoking presentations and considered discussions, including: Sam Hasan (GSK); Dale Sanders, (Intelligent Medical Objects); Stephan Reiling (Novartis); Pierre Larmande (IRD, University of Montpellier); Suman Bera (Katana Graph); Hélène Royo (Roche); Rachel Benzies (Syngenta); Kalpana Panneerselvam (EMBL-EBI); Peter Saffrey (Biomodal).

Additionally, the authors would like to acknowledge all the participants of these two workshops for the rich discussions and brainstorming that led to the central topics, and core ideas that have been vital for this paper. Organizations represented at these workshops, and not authors of this paper, were: Astellas, Astex, AstraZeneca, Boehringer-Ingelheim, Bristol-Myers Squibb, Daiichi Sankyo, Exscientia, Janssen, Pfizer, Sanofi, and Takeda.

**Disclosure Statement**

The authors report there are no competing interests to declare.